**Title**

"Elastomeric focusing enables application of hydraulic principles to solid materials in order to create micromechanical actuators with giant displacements"

**Authors**

Nate J Cira, Jason W Khoo, Mika Jain, Jack T Andraka, Morgan L Paull, Amber L Thomas, Kevin Aliado, Chad Viergever, Feiqiao Yu, Jonathan B Li, Canh T Nguyen, Michael Robles, Ismail E Araci, Stephen R Quake

**Abstract**

A continuing challenge in material science is how to create active materials in which shape changes or displacements can be generated electrically or thermally. Here we borrow principles from hydraulics, in particular that confined geometries can be used to focus expansion into large displacements, to create solid materials with amplified shape changes. Specifically, we confined an elastomeric poly(dimethylsiloxane) sheet between two more rigid layers and caused focused expansion into embossed channels by local resistive heating, resulting in a 10x greater relative displacement than the unconfined geometry. We used this effect to create electrically controlled microfluidic valves that open and close in less than 100 ms, can cycle >10,000 times, and operate with as little as 20 mW of power. We investigate this mechanism and establish design rules by varying dimensions, configurations, and materials. We show the generality of elastomeric focusing by creating additional devices where local heating and expansion are generated either wirelessly through inductive coupling or optically with a laser, allowing arbitrary and dynamic positioning of a microfluidic valve along the channels.

**Text**

Confining geometries are commonly used to realize amplifying effects in liquid based actuators. For example, a hydraulic press uses a large displacement but small force to create a small displacement but large force by Pascal's Principle, and a liquid thermometer uses a small change in volume to create a large change in displacement. Here, instead of using a liquid, we show that solid elastomers can be similarly confined to achieve disproportionately large thermal displacements and demonstrate how this mechanism can be used to create micromechanical actuators such as portable microfluidic valves.

Pneumatic, electrostatic, piezoelectric, magnetic, and thermal mechanisms[1] have been used to generate micromechanical actuation[2]. However, each mechanism has associated challenges. Small strains for piezoelectric mechanisms, the requirement for close proximity between a flexible conductor and an electrode in electrostatic mechanisms, and alignment of a mobile magnetic core with conductive coils for electromagnetic mechanisms necessitate precise device architectures and complex fabrication processes. Additionally, high voltages are required for piezoelectric and electrostatic mechanisms, while electromagnetic mechanisms require high current. These constraints have prevented devices utilizing any of these three mechanisms from scaling to the cost and size needed to put multiple actuators on a disposable device.

Pneumatic actuation has been used for macroscale actuators in soft robotics[3] and microscale actuators such as microfluidic valves[4], which are created by soft lithography and can perform complex liquid handling and enable numerous applications where precise automated liquid handling is essential such as single cell analysis[5,6] or sensitive measurements of binding interactions[7,8]. However, such pneumatic actuation is controlled by separate off-chip solenoid valves. These solenoid valves are ~1x1x1 cm and each one draws several hundred milliwatts of power, hindering development of truly portable or autonomous applications.

Efforts using thermal expansion as a means for actuation include the use of thermally deflecting materials such as shape memory alloys[9,10] or bimetals[11–13] and use of thermally expansive materials including wax[14–16], plastic[17], gas[16,18–31], liquid[23,24,28–30,32,33], hydrogels[34], elastomers[35], and composites[36,37]. Thermal deflection mechanisms require pinned boundary conditions and precise geometries created by many step fabrication processes and often large footprints to leverage small shape changes due to a change in temperature into relevant displacements. Thermally expansive materials are often chosen to have large thermal expansion coefficients, such that the volume change from heating can be used directly for actuation without amplification. However, these must still be patterned at each actuation location, and containing hot pressurized gas or liquid within flexible yet impermeable boundaries is a challenge. All of these previous thermal actuation mechanisms are sensitive to the ambient temperature of the operating environment. Our approach further simplifies fabrication by eliminating the need to locally pattern the thermally expansive component, instead using a continuous solid sheet of poly(dimethylsiloxane) (PDMS) directly as the actuator. This avoids the challenges of containing liquids and gases, and makes the actuation insensitive to ambient temperature.

Though high for a solid, the modest thermal expansion coefficient of PDMS ($3\times10^{-4}$/°C)[38] means that heating PDMS is ordinarily insufficient to generate displacements large enough to close typical microfluidic channels. For example, a 150 µm thick layer of PDMS heated by 20 °C only increases 0.9 µm in height. However, since PDMS is essentially incompressible at reasonable pressures[39] the small volumetric expansion from a larger area can be focused into a large displacement in a smaller area with a confining geometry. We accomplish this by placing a PDMS layer between two more rigid layers, a glass base and a polystyrene (PS) top (figure 1A, supplementary figure 1). We emboss channels into the polystyrene top, and pattern clear resistive heating elements in indium tin oxide (ITO) on the glass layer. When voltage is applied across the ITO resistor, Joule heating and thermal expansion of the PDMS occur. Since the PDMS is confined by the rigid layers on the top and the bottom and the surrounding PDMS on the sides, this expansion is forced into the embossed channel, with a displacement sufficient to close the channel, forming a functional valve (figure 1B, supplemental video 1, for full fabrication details see materials and methods). Using this method, we have been able to close channels over 30 µm tall with PDMS layers as thin as 160 µm. Simulations show that this represents a 12-fold increase in the channel height that would be possible to close with conventional thermal expansion, ie: with the same temperature profile but without rigid confinement. (figure 1C)

Thermal expansion results in a total volume change, $\Delta V$ of $\Delta V = 3\alpha \iiint \Delta T(x, y, z) dx\, dy\, dz$, where $\alpha$ is the linear thermal expansion coefficient, and the integral is taken over the entire volume of heated material. This expansion could close a length of channel $L = \frac{\Delta V}{A}$, where $A$ is the cross-sectional area of the channel if the entire volume from expansion redistributed itself into the channel, as it would with a hydraulic liquid, capable of redistributing into any new shape. However, since the expanding material here is an elastomeric solid with fixed boundary conditions on the top and bottom (the three layers are permanently bonded together), it is unable to redistribute as efficiently as a liquid, and the effective length of channel closed is always smaller than this maximum. The remaining volume expansion deflects the boundary substrates apart. Given the complexity of this geometry we used experimental observations and numerical simulations to optimize device configurations and understand the limits of elastomeric focusing.

In support of the proposed mechanism, insufficient stiffness of the boundary substrates prevents the crucial confinement effect and does not form a functional valve. Devices where we replaced the top polystyrene layer with PDMS (4000x lower Young's modulus) failed to actuate at any applied power. Simulations also indicate that the top layer must be a few orders of magnitude more rigid than the middle PDMS layer for effective focusing (figure 1C). A view of the lateral displacement from numerical simulations supports the redistribution of the elastomer towards the channel from an area wider than the channel itself (figure 1D).

As a demonstration of the utility of elastomer focused actuators, we created a series of microfluidic valves with a variety of performance and actuation properties, including both electrical and optical means.

We fabricated 72 different valve geometries with varying channel heights, channel widths, PDMS thicknesses, and PDMS mixing ratios, and we actuated each of these at seven different powers. This dataset enabled us to create a set of design rules which can be used to optimize valve performance for desired metrics. For example, we were able to create valves which require as little as 18 mW of power to close, and valves that open and close with speeds comparable to or faster than the frame rate of the camera (41 ms). The tallest channel heights we fabricated were around 30 μm, but there is no inherent theoretical maximum channel height. Increasing channel heights at the same PDMS thickness require linearly more power to close (figure 2D), which can be understood heuristically since thermal expansion is linear with temperature which is approximately linear with power.

Figure 2A indicates the length of a closed section for one valve over time at several powers. From videos of the valve opening and closing at each power we extracted a minimum closing time, equilibrium closed valve length, and minimum opening time. Increasing power decreases closing time, increases opening time (figure 2B) and increases valve length (figure 2C). Applying a constant power was useful for characterizing performance, but the power can be easily controlled to obtain the best combinations of characteristics. For example, a 150 ms pulse of 120 mW followed by pulse width modulation with a 32% duty cycle (effectively 38 mW) created a 105 μm valve which both opens (41 ms) and closes (125 ms) rapidly. (red trace, figure 2A).

The optimal PDMS thickness varied, and was typically on the order of the channel width (figure 2E). Mixing ratio changes PDMS elasticity[40,41] which has an effect on valve performance – stiffer layers open faster, and require more time and power to close, especially for narrow channels (supplementary figure 2). To quantify any variability in valve performance, we measured the minimum actuation powers and report means ± standard deviations of a single valve tested five times (47 ± 1.5 mW), five valves on the same chip (44 ± 2.4 mW), and six valves across six different chips (38 ± 4.8 mW) (supplementary figure 3, A-C). For the complete dataset on all the valves tested, see the appendix.

By testing valves with various spacings, we determined the minimum valve to valve spacing along a channel to be as low as 388 μm in unoptimized designs (supplementary figure 4), and we anticipate that passive or active heat sinks could improve packing density. Valve length had negligible rate-independent hysteresis as power was varied at minute timescales (supplementary figure 5). However, leaving the valves in the powered state over hours resulted in an increase in the power required to close the valve, and valve recovery happened over a similar timescale (supplementary figure 6), which we hypothesize is due to viscoelastic creep of the PDMS. Apart from the increase in actuation power due to creep, valves were fully reversible, operating consistently across over 10,000 actuation cycles without noticeable change in performance.

We experimentally measured the temperature distribution around the valve by injecting waxes of known melting points into the channel and applying different powers and measuring the length of melted wax (supplementary figure 7). These results can be partially understood in the context of a simplified model of the powered valve as a constant heat flux from a point source, diffusing radially in three dimensions. This results in a temperature change $\Delta T = \frac{P}{4\pi k r}$, where $P$ is the power, $k$ is the thermal conductivity, and $r$ is the distance from the source, consistent with the linear increase in the length of melted wax as power was increased. Since the channel is offset from the resistor by the PDMS thickness, $h$, the model can be further modified to $\Delta T = \frac{P}{4\pi k \sqrt{x^2+h^2}}$, where $x$ is the distance along the channel, which helps capture the observation that some power is required before any wax melts. However, a quantitative fit with this simple model is confounded by the finite shape and composite nature of the device. Based on our wax measurements and depending on valve geometry, the temperature of the liquid in the channel adjacent to a closed valve will increase by 15 C or more (supplementary figure 7). Active or passive heat sinks and adjustments of the material thermal conductivities or layer dimensions are expected to minimize heating for sensitive applications.

Unlike all other thermally actuated mechanisms that rely on spatially patterned expansive materials, these valves require local heating to operate but are insensitive to global temperature variation, since global temperature shifts simply change the total thickness of the PDMS layer. In agreement with this understanding, we tested valves at 6 C, 23 C, and 50 C, and found comparable performance (Figure 2F).

We took most of our measurements on channels filled with air, since the higher difference in refractive index simplified visualization and analysis. When channels were filled with liquid, similar trends held, with valve length and valve opening time within around 15% of the measured values with air but valve closing took about 75% longer, possibly due to increased viscosity of the liquid in the thinning film between layers as the valve closes (supplementary figure 8).

We noticed that rapid actuation of the valves sometimes caused bubbles to appear in the channel when loaded with liquid (supplementary video 2). The bubbles formed upon the valve reopening. We found that we could overcome bubble formation by 1) maintaining the device in a vacuum (0.1 barr for 2 hours) prior to use, with this protective effect lasting at least 1 day, or 2) maintaining back pressure (5 psi) on the liquids during actuation (supplementary video 3).

Elastomeric focusing allows a diversity of fabrication options. As long as the top and bottom layers are sufficiently more rigid than the elastomer, multiple different configurations and material choices will create functional valves. For example, we made devices where the channels were patterned in the top of the PDMS layer, and the polystyrene layer was flat (figure 3A, supplementary video 4) which performed comparably to placing the channels in the top layer (supplementary figure 9). We also ordered and used commercially manufactured printed circuit boards (PCBs) with both resistive alloy or printed carbon resistors to replace the ITO/glass base (figure 3A), supplementary videos 5&6). The low cost of PCBs suggests that scaled manufacturing would be inexpensive, especially when paired with an injection molded plastic component. It is interesting to note that to reach a given temperature profile, power is inversely proportional to thermal conductivity, so the actuation power is predicted to decrease when we switch glass ($k$ = 1.1 $\frac{W}{m \cdot K}$)[42] with FR4 fiberglass ($k$ = 0.29 $\frac{W}{m \cdot K}$)[43]. While valves on PCB substrates opened and closed effectively, we did not observe this predicted drop in actuation power (supplementary figure 10). Thermal imaging suggested this was because heat was carried away by the copper traces (supplementary figure 11) which have orders of magnitude higher thermal conductivity than FR4 ($k_{Cu}$ = 385 $\frac{W}{m \cdot K}$)[43]. Minimizing height and width of these traces, especially near the valves is predicted to decrease actuation power in future designs. Additional boosts in efficiency might also be realized by adding lateral rigid confinement to the heated elastomer.

    We explored alternative ways to power the valves. Inductive coupling is well established as a means for wireless power transfer, and has been used to power drug delivery devices internal to the body[44,45]. As a proof of concept, we showed it was possible to power the valves wirelessly through inductive coupling (figure 3B). Local heat can also be supplied by means other than joule heating. We made a photoabsorptive layer by spiking carbon black into the PDMS layer (1% by mass, 210 µm height) and replacing the bottom ITO/Glass layer with a simple glass slide. This configuration enabled valve actuation via laser (100 mW, 532 nm), and the position of the valve was no longer fixed, but could be created and moved dynamically anywhere along the channel by moving the beam position (figure 3C, supplementary video 7). We also made a laser actuated device which had patterned photoabsorptive portions and showed that valves were only created when the beam position intersected these patterned regions (figure 3D, supplementary video 8).

    It is worth noting that this mechanism of valve actuation allows analog control of valve closure length (Figure 2C), which could be useful for high precision metering of liquids, or control of multiple fluid paths with a single active valve. Even before complete closure occurs, constriction of the valve could be used to tune the fluidic resistance of channels, or to create a sieve valve[46] at any valve location. These valves can also be integrated to replace solenoid

valves, augmenting the portability of existing pneumatic systems. Electrical connections to the chips make possible additional functionalities such as thermal cycling and temperature control for temperature sensitive reactions, electrophoresis for separations, impedance cytometry, electrical detection of DNA, and electrolysis to create localized pressure sources.

Easily controlled portable valves have potential applications in a wide range of devices. Handheld integrated devices could be created to execute the multiple liquid handling steps required to get from sample to result in applications such as food and water safety, outbreak monitoring, and clinical and personal diagnostics. Low power consumption valves could see uses in wearable or implantable devices, or could be integrated for the motion of mini-robots. To demonstrate the portability of these valves, we created a handheld battery powered valve controller capable of opening and closing a valve according to a preprogrammed sequence (supplementary video 9). Finally, elastomeric focusing does not solely apply to generating amplified displacements from thermal expansion of PDMS; displacements from piezoelectric actuators, liquid crystal elastomers[47–49], or electroactive polymers[50,51] could also be focused into larger gains using this mechanism.

**Acknowledgements**

We gratefully acknowledge contributions from Karen Lee, Nathan Orloff, and Jotthe Kannappan. Cleanroom support was generously provided by Adam White. NJC was supported by an NSF GRFP and Siebel Scholar Fellowships. FBY was supported by NSF GRFP and Stanford Graduate Research Fellowships, and MLP was supported by an NSF GRFP Fellowship. Funding for this work was generously provided by the TomKat Center for Sustainable Energy and by the Stanford Clinical and Translational Science Award (CTSA) to Spectrum (UL1 TR001085). The CTSA program is led by the National Center for Advancing Translational

**Figures and captions:**

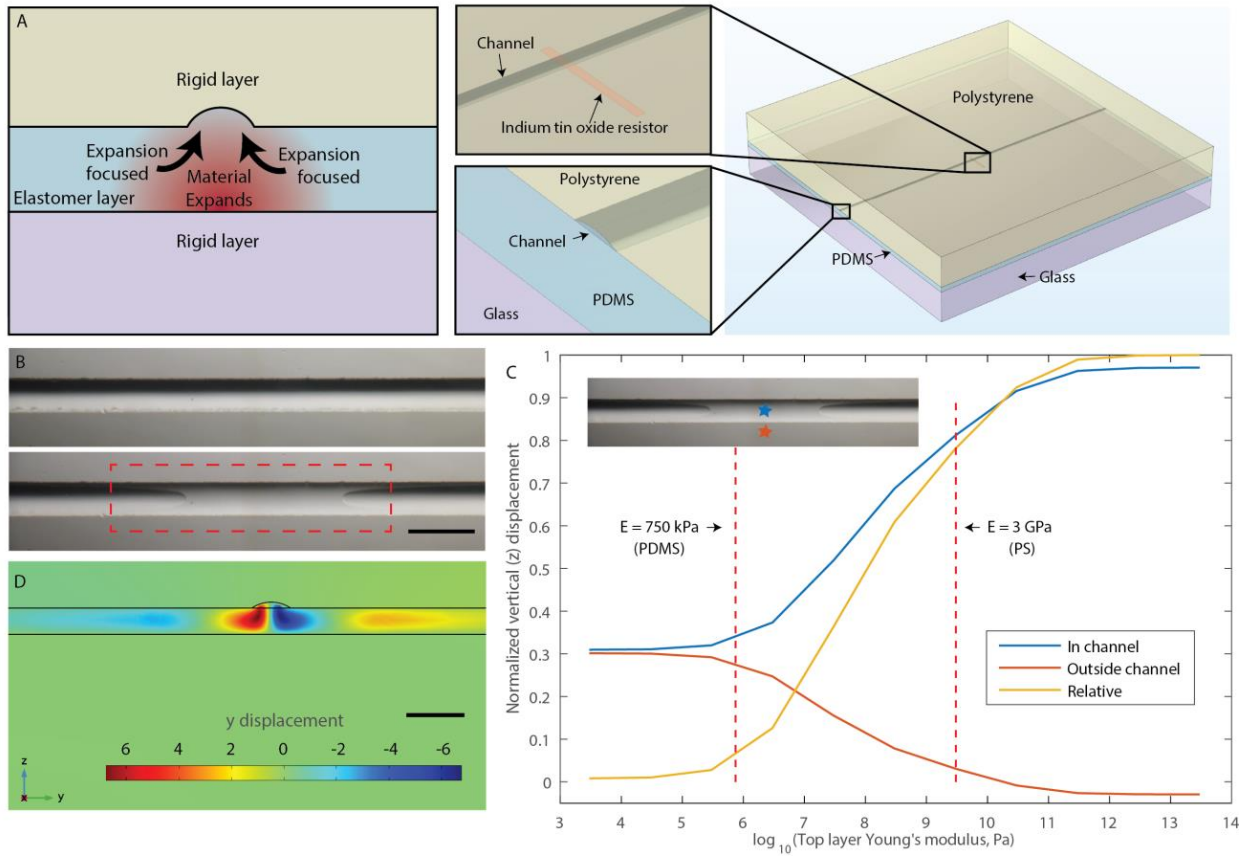

**Figure 1.** Mechanism. A) A schematic showing the device geometry. The device has three layers, a bottom glass layer, a middle PDMS layer, and a top polystyrene layer. Resistive heating elements are patterned in indium tin oxide on the top surface of the glass, and channels are embossed into the polystyrene. Color is for visualization only. B) Top-down photographs of a microfluidic valve in the open (top) and closed (bottom) states. C) Results from numerical simulations show the effect of top layer rigidity on confinement. The solid red line indicates the vertical displacement of the top of the PDMS outside the channel (position indicated by red star on inset). The blue line shows the maximum displacement of the top of the PDMS inside the channel (position indicated by blue star on inset). The gold line is the difference between the two, and indicates the effective height of channel that can be closed. Traces are normalized to the maximum displacement, and zero is the initial height before power is applied. The dashed red lines indicate Young's modulus for PDMS and polystyrene. The top layer must be substantially more rigid to obtain effective focusing. D) A plot of the cross section of the channel showing y-displacement. In the vicinity of the channel, the PDMS redistributes towards the channel, focusing displacement (units are µm). Both scale bars are 200 µm.

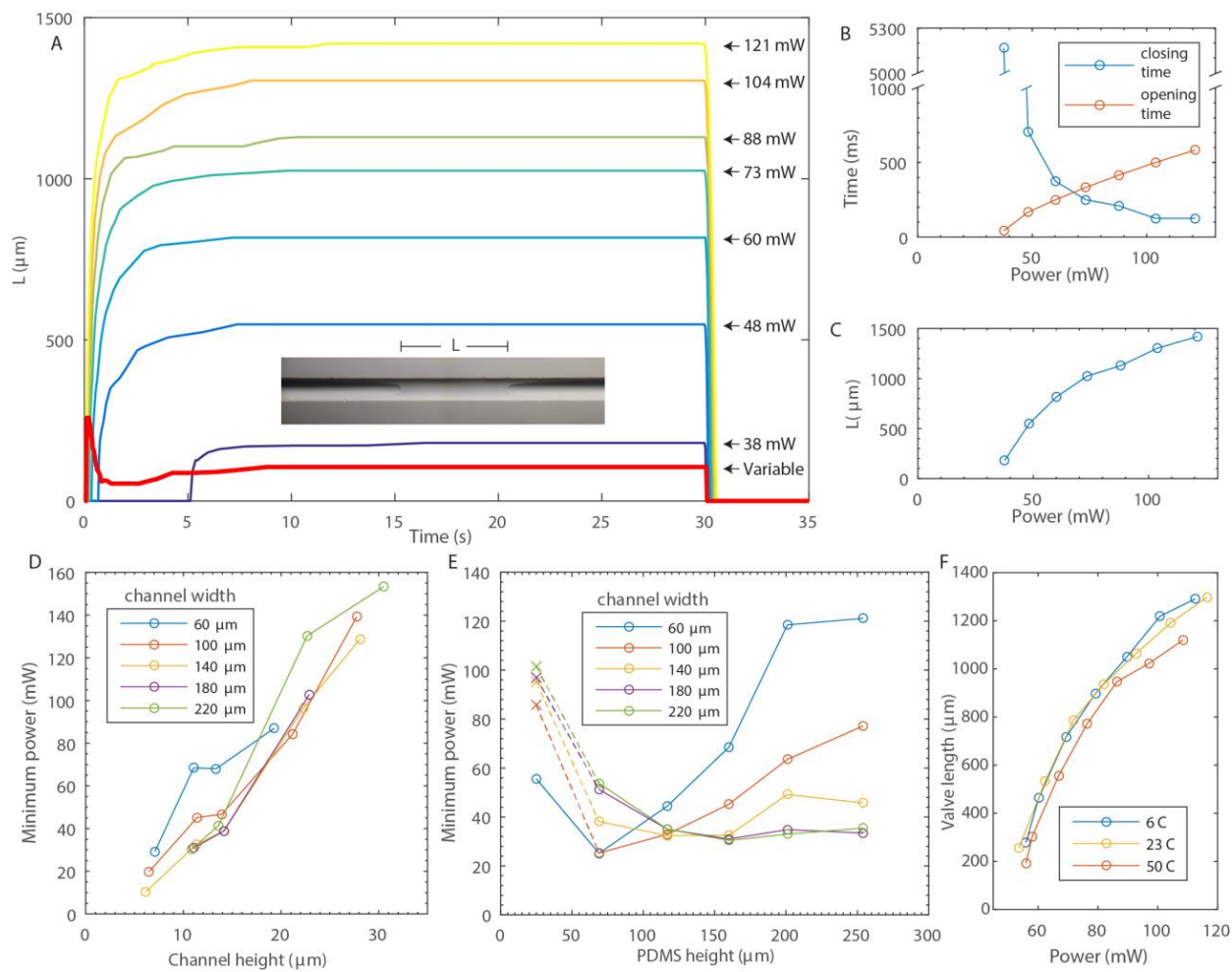

**Figure 2.** Valve characteristics. A) Seven different powers were applied to a valve for 30 s. Length of the closed section is plotted over time for one device with PDMS layer thickness of 117 µm, channel height of 11.4 µm, and channel width of 100 µm. Power for the red trace is 120 mW for 150 ms, then 38 mW. B) The opening (red) and closing (blue) times of the valve in A. Opening time was the difference between when the power was switched on and when the channel was blocked, and closing time was the difference between when the power was removed and when the channel reopened. C) Length of the closed section is plotted against power for the valve in A. D) The minimum power required to close the valve is plotted against the channel height, and increased approximately linearly for 5 different channel widths when all other parameters were held constant (PDMS layer height = 160 µm) E) The minimum power required to close the valve is plotted against the PDMS layer height for different channel widths. Dotted lines connect the data points to x's, which denote failure before closure. Optimal PDMS layer thicknesses seem to exist for minimizing power and increase with increasing channel width. Channel height in E was around 11 µm for all widths. F) Valve length vs power is consistent in different background temperatures.

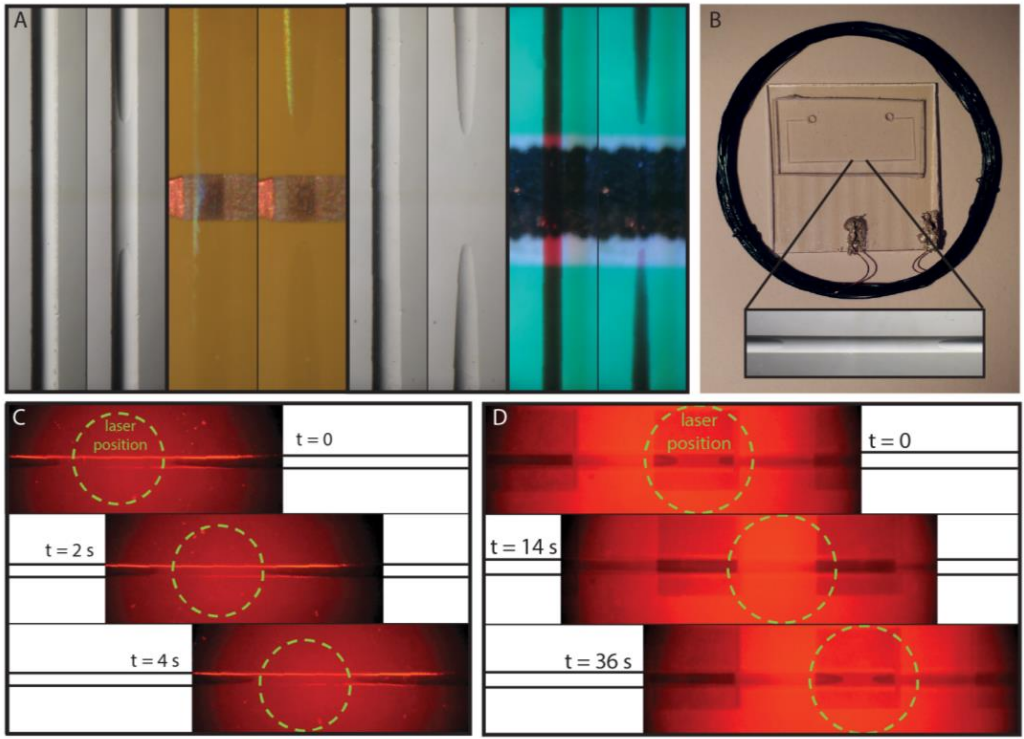

**Figure 3.** Permutations and applications. A) In addition to the standard valve described in the main text (left), we created additional permutations. We created valves where the channels were located in the PDMS layer and the PS layer was flat (second from right), valves where the resistors were patterned on printed circuit boards in resistive alloy (second from left), and printed carbon (right). To improve image contrast we filled the channel above the printed carbon resistor with red food coloring. B) We attached an inductive coil to a chip, and demonstrated that this addition allows the valves to be wirelessly activated. C) We created a photoabsorptive layer by doping the PDMS with carbon black. By shining a 100 mW green laser at this device we were able to achieve local heating which created a valve. Since the location of the laser can be moved around the channel, this valve is dynamic, and can even turn corners (Supplementary video 7). The green dashed line indicates the FWHM position of the beam. D) We also patterned the doped PDMS, which created permissible and prohibited valve locations, since the doped areas absorb light and heat up, while the clear positions allow the beam to pass without substantial heating.

**Supplemental:**

**Materials and Methods**

*Chip fabrication*

The microfluidic chips consisted of three layers that were fabricated separately and bonded together.

*ITO/Glass layer*

The glass/ITO layer was created by blocking and wet etching. We purchased glass slides that were precoated with ITO from two different suppliers (Sigma Aldrich 703192, surface resistivity 8-12 Ohms/sq and Techninstro TIX001, surface resistivity ~10 Ohm/sq). We spun positive photoresist (Megaposit™ SPR 220.7) on the ITO coated surface and used photolithography to pattern the photoresist, according to the manufacturer's guidelines. After development (Microposit ® MF 319) we placed the slide on a hotplate for an additional 90 seconds at 90 C to help solidify the resist. We then used a 1:1 mixture of 37% hydrochloric acid to deionized water with agitation to etch away the ITO coating from areas not covered in resist. We etched until the exposed regions were no longer conductive (typically 70 minutes). When etching was complete we removed the photoresist with acetone and isopropanol. For resistors, we used rectangles of 50 x 700 µm, and for conductive paths we used 1 mm wide traces. Alternatively, we also found it was possible to pattern the ITO using a programmable ultraviolet laser. Use of ITO/Glass enabled transparent devices, but opaque conductors can also be used.

*PDMS Layer*

We treated clean silicon wafers with chlorotrimethylsilane for 10 minutes to reduce adhesion, then spun PDMS (GE: Momentive RTV615) onto the wafer and cured it at 80 C for a minimum of one hour. Layer thickness was controlled by the spinning speed, and various mixing ratios (from 1:5 to 1:20, B:A mass:mass), and layer thicknesses (25 µm - 254 µm) were used, as described in the text. Dow Corning's Sylgard 184 PDMS also resulted in functional valves.

*Polystyrene layer*

Channels were patterned in a sheet of polystyrene using hot embossing off of silicon/photoresist or epoxy molds.

We created molds using photolithography on silicon wafers. Wafer were first treated with hexamethyldisilazane (HMDS) adhesion promoter, then positive photoresist (Megaposit™ SPR 220.7) was spun to the desired height on a spin coater (6 µm to 30 µm). Exposure and development (Microposit ® MF 319) were carried out to the manufacturer's recommendations. We hard baked the patterned wafer on a hot plate by ramping from 65 C to 190 C at 10 C/hour, holding for 2.5 hours at 190 C, then allowing the hotplate to cool. During this process the photoresist reflows, resulting in a rounded cross section which is critical to obtaining complete valve closure.

We used a hot embossing method inspired by Vasiliy *et al*[52]. This method consists of making a mold and clamping a thermoplastic to a mold and baking it. Unlike Vasiliy *et al*, we did not use PDMS molds, since these were not sufficiently flat for our purposes.

We found we were able to emboss directly off the silicon wafer with patterned photoresist if the features were less than 10 µm in height. For features taller than 10 µm we found that the SPR photoresist features occasionally broke from the silicon wafer during embossing. To overcome this, following Young et al.[53], we created epoxy molds to use for hot embossing. This was accomplished by casting PDMS off a patterned wafer then casting an epoxy mold off the PDMS. We found Axson EC-415 epoxy cured to the manufacturer's recommendations worked well. We did not notice deterioration of these molds, despite using some for more than 20 embossing cycles.

The hot embossing was carried out with the same protocol regardless of mold type. We clamped polystyrene sheets (we used flat portions from petri dishes ~1 mm thick as the PS substrate) between the mold and a flat glass plate using binder clips which applied around 10 lbs of total force on pieces typically around one square inch. This assembly was placed in a convection oven at 150 C for 15 minutes, then removed and allowed to cool to room temperature on the benchtop prior to demolding. The embossed PS chip was trimmed to size, if necessary, and through holes were drilled when access ports to the channels were required. We found it was important to bake the polystyrene at 80 C overnight before embossing to avoid bubbles. The process was not sensitive to variations on baking times or clip type. We note that the process of producing these plastic components could be scaled up with injection molding.

*Bonding embossed polystyrene layer to spun PDMS*

The polystyrene layer was bonded to the PDMS layer using APTES/GPTS treatment as in Tang and Lee.[54] The PDMS (still attached to the wafer) and polystyrene layers were treated with oxygen plasma in an Anatech Ltd SP100 Plasma System with machine settings of 35 seconds exposure at 80 watts.

Next, the PDMS-coated silicon wafer was coated with 2% (3-Aminopropyl)triethoxysilane (APTES) in water solution for 20 minutes. Concurrently, the polystyrene chip was treated with a 2% (3-Glycidyloxypropyl)trimethoxysilane (GPTS) in water solution for 20 minutes. Both substrates were then rinsed with deionized water and blown dry. The treated surfaces were then placed together to form a bond. The assembly was then placed in an oven at 80 C overnight to complete bonding. The bonded assembly was then removed from the wafer by slicing through the PDMS around the polystyrene chip with a scalpel and peeling the polystyrene chip and the area of the PDMS bonded to it from the wafer.

*Bonding ITO/Glass layer to combined PDMS/polystyrene layer*

The final step is to bond the polystyrene/PDMS chip assembly to the patterned glass/ITO. This was done again by treatment with oxygen plasma in an Anatech Ltd SP100 Plasma System with machine settings of 35 seconds exposure at 80 watts.

The two plasma oxidized surfaces are aligned and placed in contact with each other, then placed in an oven at 80 C for 1 hour to create a finished chip.

*Chip permutations*

We also created chips where the channels were patterned in the elastomer layer adjacent to a flat polystyrene layer. These were created in a similar fashion to the configuration described above, except the elastomer was spun and cured onto a wafer with the desired channel patterns, then plasma bonded to the etched ITO/Glass, then a flat polystyrene layer was bonded to the elastomer using APTES/GPTS treatment as described above.

The PCB designs were fabricated by the same methodology as the standard chip, except that the PDMS was spun and cured directly onto the PCB, which effectively eliminated one bonding step. Note that this is also possible with ITO/glass, but we used 25 x 25 mm ITO/Glass slides and edge effects from spinning resulted in an insufficiently flat layer. Larger Glass/ITO substrates could also use this shortcut. PCBs themselves were ordered from commercial suppliers. We ordered printed carbon resistors on CEM-1 boards from Gesp Technologies (Dalian) Co, and alloy resistors (Nickel Chromium, Ticer) on FR4 fiberglass boards from Shipco Circuits. Typical resistor dimensions were 250 x 1000 µm for the printed carbon resistors and 200 x 200 µm for the alloy resistors.

*Chip characterization*

We characterized chips by filming the valves while applying a 30 s pulse of power. We used a programmable Arduino board and a transistor to control the timing of the power pulse and a power supply to control the applied voltage. We recorded valve responses using a Canon Powershot G15 camera fixed to a Nikon SMZ1500 stereoscope. We used the image analysis program ImageJ to step through frames, recording the frames at which power was applied and removed, when the valve opened and closed, and measuring the length of the closed valve. This methodology gave us a minimum time resolution determined by the frame rate of the camera (24 fps = 41 ms time steps). We report the power as only the power delivered to the heating resistor, or, the total power dissipated by the circuit minus the power dissipation in other parts of the circuit (transistor, contact resistance, and dissipation in the wires/legs of the devices)

Thermal images, as in supplementary figure 11, were acquired with a Seek Thermal infrared camera mounted on an iPhone 6.

*Wireless chip*

We built a wirelessly powered chip by attaching a receiver coil (30 turns, 4 cm diameter) to the chip with conductive silver epoxy. We then used a paired transmitter coil (20 turns, 4 cm diameter) powered by a function generator (Stanford research systems, DS335, 700 kHz, 10 Vpp, square wave) to drive current through the resistor by inductive coupling.

*Laser powered chips*

We created a laser powered chip by doping the PDMS layer with carbon black. In one device we spun a 1% carbon black PDMS layer to 210 µm on a 50 x 75 mm glass slide, and bonded an embossed PS layer to the black PDMS.

In the second device, we spun clear PDMS on a wafer patterned with square posts (500 x 500 µm wide x 40 µm tall with 500 µm spacing) to a height of 157 µm. We bonded an embossed PS layer to this spun and cured PDMS layer, then removed the assembly from the wafer. We then pressed this assembly into uncured 1.5% carbon black PDMS on a glass slide, such that the post locations (wells in the clear PDMS) were filled with black PDMS and the spaces between these features were transparent. The device was then cured for 24 hours at 45 C.

Both of these devices were powered with a 100 mW green laser.

*Valve controller*

We created a valve controller by using a programmable Arduino Micro board to control a transistor (Fairchild: 2N3904BU) which controlled whether power was delivered to the valve by a 9 V battery. Power level was controlled by pulse width modulation. We attached the circuit to the back of a phone case and used a Samsung Galaxy S4 and acrylic lens (AIXIZ: AIX-LENS-123) with an LED backlight to visualize the valve opening and closing. The chip interfaced with the circuit through a card edge connector (Sullins: EBC10DCWN) and ribbon cable.

*Numerical modeling*

We used COMSOL Multiphysics software to generate an *in silico* model of the valves to rapidly test permutations, gain intuition, and see trends without the need to make and test every physical design. COMSOL captures device geometry, material properties, and initial conditions, discretizes the model, then numerically solves the system.

Each layer was 10 x 10 mm wide, and the top (PS) and bottom (Glass) layers were 1 mm thick. The PDMS layer height was a tunable parameter, as were the channel height and width. The resistive heating element was created as a 2-dimensional rectangle on the top of the glass surface. We assigned different material properties to the different layers of our device. We used the values in table 1-3 for the material properties of the layers.

| | | |
|---|---|---|
| Coefficient of thermal expansion | alpha | 3e-4[1/K] |
| Heat capacity at constant pressure | Cp | 1460[J/(kg*K)] |
| Density | rho | 970[kg/m^3] |
| Thermal conductivity | k | 0.16[W/(m*K)] |
| Young's modulus | E | 750000[Pa] |
| Poisson's ratio | nu | 0.49 |

Table 1: PDMS material properties.

| | | |
|---|---|---|
| Coefficient of thermal expansion | alpha | 70e-6[1/K] |
| Heat capacity at constant pressure | Cp | 1420[J/(kg*K)] |
| Density | rho | 1190[kg/m^3] |
| Thermal conductivity | k | 0.19[W/(m*K)] |
| Young's modulus | E | 3e9[Pa] |
| Poisson's ratio | nu | 0.40 |

Table 2: PS material properties.

| | | |
|---|---|---|
| Coefficient of thermal expansion | alpha | 0.55e-6[1/K] |
| Heat capacity at constant pressure | Cp | 703[J/(kg*K)] |
| Density | rho | 2203[kg/m^3] |
| Thermal conductivity | k | 1.38[W/(m*K)] |
| Young's modulus | E | 73.1e9[Pa] |
| Poisson's ratio | nu | 0.17 |

Table 3: Glass material properties.

We used the following boundary and initial conditions: An initial temperature of 293.15 K, a convective heat flux on all outside borders of the device of $70 \frac{W}{m^2 \cdot K}$, a constant heat flux to the resistor (tunable parameter), and fixed the outer edges of the glass layer (perimeter of the 10 x 10 mm square, but not the top or bottom of this layer) in space.

We used the automatic meshing function to discretize the system and specified a region of denser mesh near the valve. We relied on steady state solutions to the system, though it was also possible to generate time dependent solutions.

We note that one aspect of the model is not physical – despite our efforts we were not able to implement contact between the top and bottom surfaces of the channel. That is, if enough power was applied, the PDMS "floor" of the valve could go through the PS "ceiling", an apparent material intersection. Also, the model does not account for any adhesive surface forces as the valve closes. Despite these shortcomings, we were still able to gain insights into the system behavior.

**Supplemental Figures:**

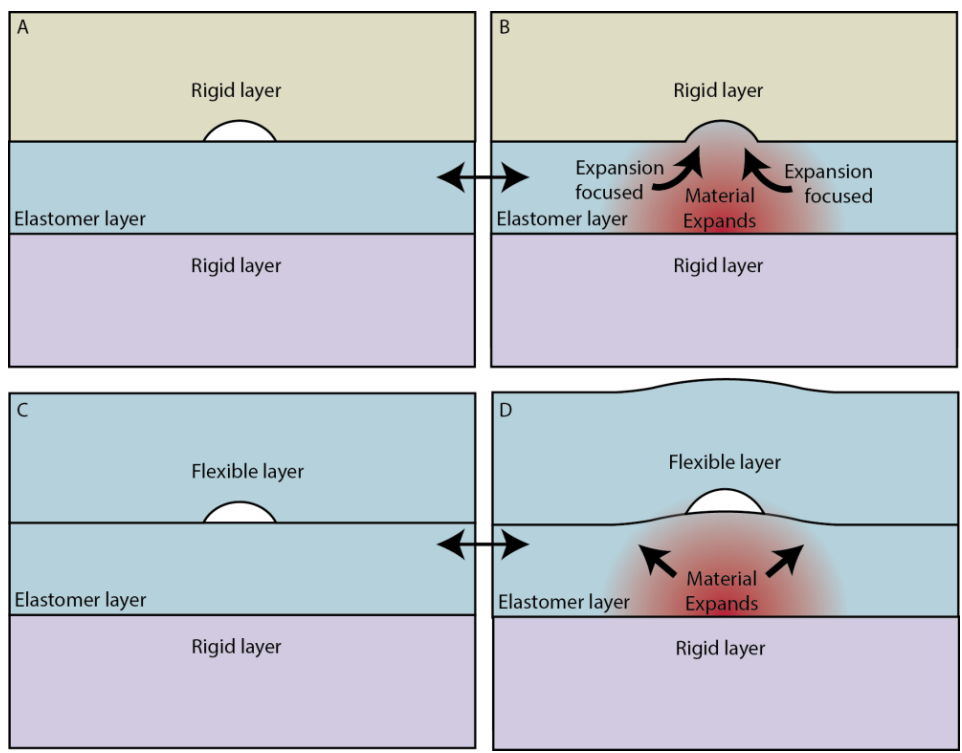

**Supplementary figure 1.** Schematic of mechanism. Elastomeric focusing (A&B) relies on confining an elastomer with more rigid materials. The rigid layers cause expansion of the middle layer to be focused into the channel, closing it (B). This is in contrast to heating the same configuration without rigid confinement (C&D). Here the flexible top does not constrain the direction of expansion, and the channel does not close.

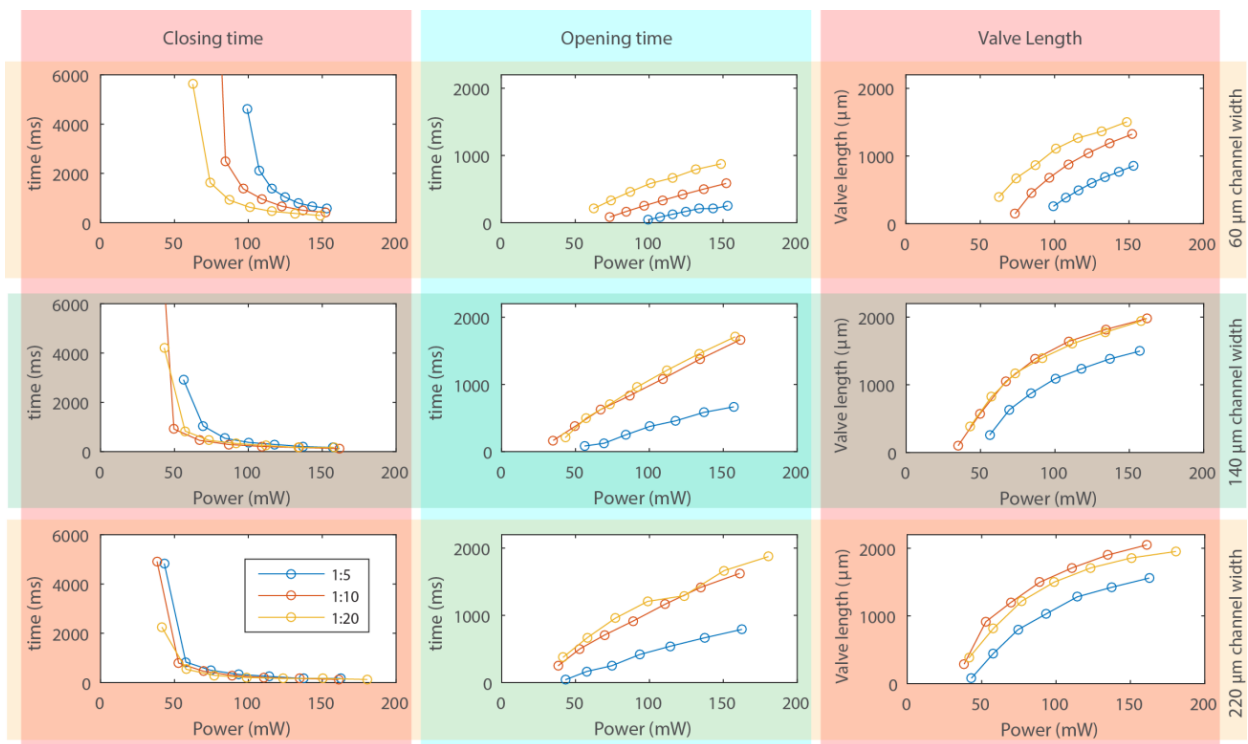

**Supplementary figure 2.** PDMS Mixing ratio (crosslinking agent:base key shown in lower left). Increasing the ratio of crosslinker to base ratio increases the PDMS stiffness. Here we display results at three different PDMS mixing ratios for three different channel widths. Stiffer PDMS required more power and time to close, but opened faster. This effect is more pronounced in narrower channels. All other parameters were held constant in this experiment (10.87-11.32 µm channel height depending on width, 160 µm PDMS thickness). For data on two additional widths see the appendix.

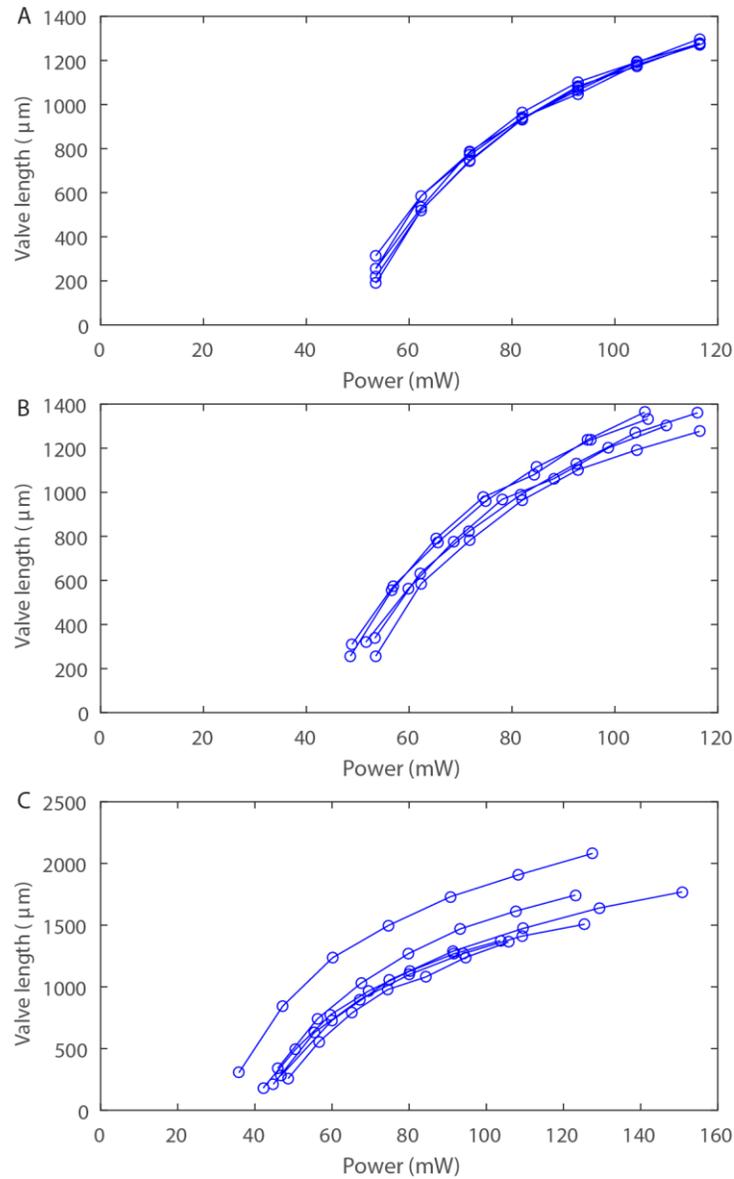

**Supplementary figure 3.** Replicates. A) We measured the length vs power relationship for five repeated measurements on a single valve B) We measured the length vs power relationship for single measurements on five different valves on the same chip C) We measured the length vs power relationship for single measurements on six different valves on 6 different chips. All valves were fabricated to the same target specifications (11.27 µm channel height, 100 µm channel width, 160 µm PDMS thickness, and 1:10 PDMS mixing ratio). The measurements in C give an indication of the expected variation due to fabrication differences.

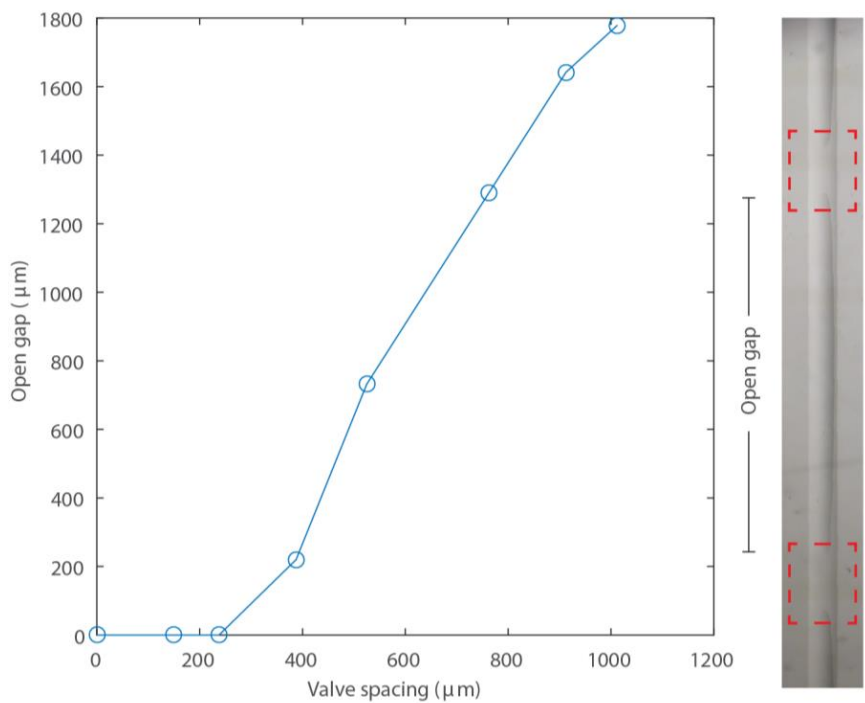

**Supplementary figure 4.** Minimum valve spacing. We created a device (PDMS thickness of 160 µm, channel height of 11.8 µm, channel width of 100 µm) with resistors spaced along the channel to obtain valves separated by differential distances. We closed the valves two at a time to determine the minimum spacing. To determine spacing, we measured the length of the open section between two closed valves at equilibrium. We reasoned if there was an opening between the two valves, a third valve could have been created in this space which could act to close the channel independently at that location. The effective valve spacing is then half the distance between the two original valves. Valve locations and the open gap from an example image are displayed on the right. Here we found that we were able to maintain a stable air gap with two valves spaced 775 µm apart, but not below 475 µm apart, regardless of applied power, so the minimum valve spacing is between 238 µm and 388 µm, at most 388 µm for this device.

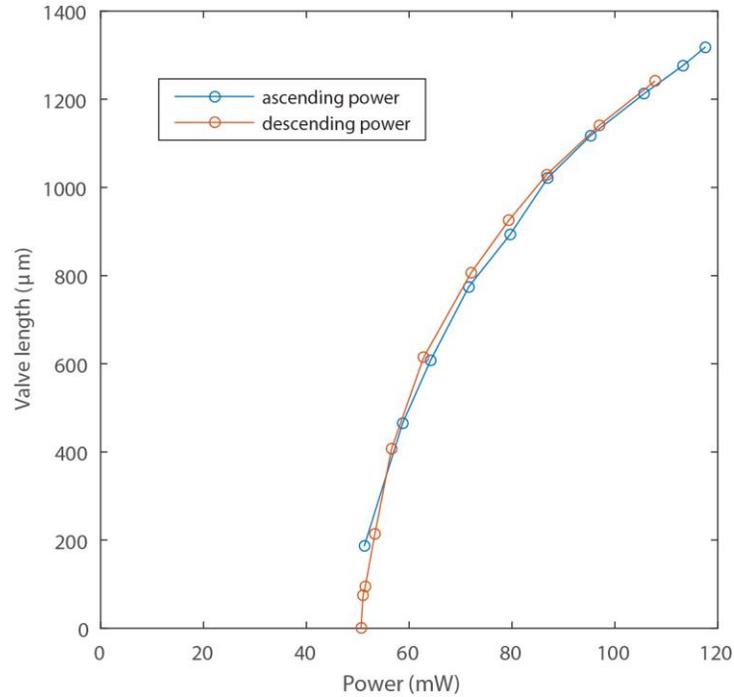

**Supplementary figure 5.** Rate independent hysteresis. We increased the power starting at the minimum valve closing voltage, pausing for 45 seconds at each measurement point for the valve to reach an equilibrium length. Then we ramped the power down, again pausing 45 seconds at each measurement point. The length vs power measurements of the ascending and descending curves show little difference, indicating minimal rate-independent hysteresis. PDMS thickness was 160 µm, channel height was 11.27 µm, and channel width was 100 µm.

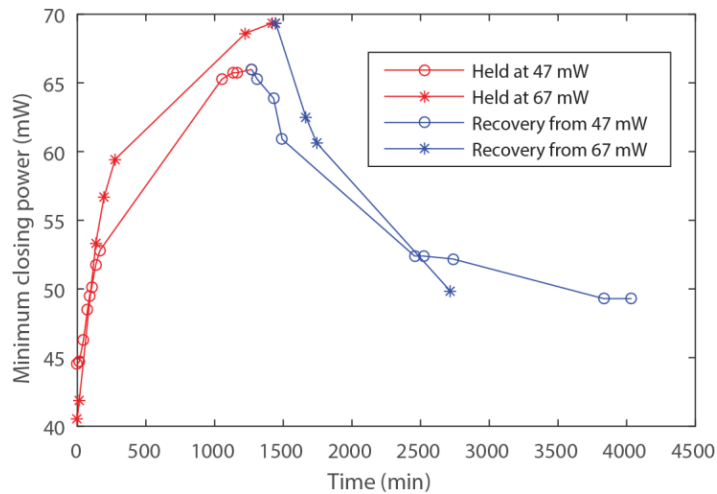

**Supplementary figure 6.** Effect of long-term valve closure. We maintained constant power to a valve for a period of one day, pausing to briefly measure the increase in minimum actuation power. We then then removed the power and allowed the valve to recover, again pausing briefly to measure the decrease in minimum actuation power. PDMS thickness was 160 µm, channel height was 11.27 µm, and channel width was 100 µm.

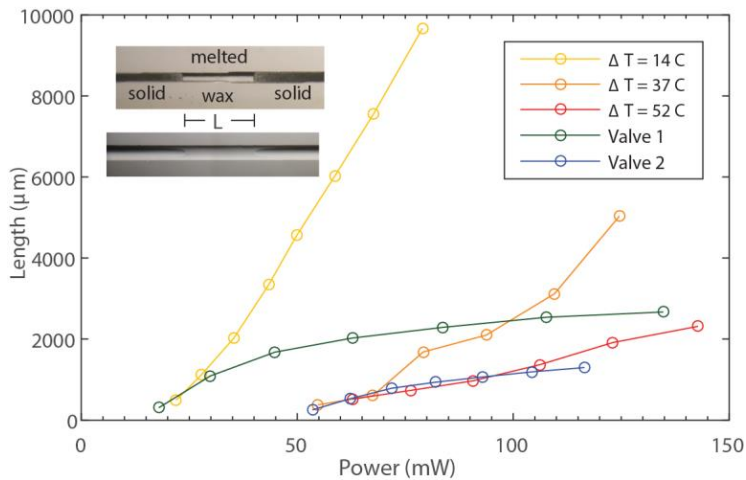

**Supplementary figure 7.** Temperature adjacent to closed valve. We injected 3 different waxes with 3 different melting points into three different channels and allowed them to solidify. We then applied different powers and measured the length of wax that liquefied within the channel. By comparing the valve lengths to the length of melted wax at different powers we can determine the temperature increase in fluid immediately adjacent to a closed valve. Here we overlay the length of two different valves. Upon closing, the liquid adjacent to the lower power valve experiences an increase in temperature of less than 14 C, while the liquid adjacent to the valve that closes at higher power experiences a temperature shift closer to 40 C. Valve parameters are 6.15 µm channel height, 140 µm channel width, 160 µm PDMS thickness, 1:10 PDMS mixing ratio, and 11.27 µm channel height, 100 µm channel width, 160 µm PDMS thickness, and 1:10 PDMS mixing ratio, respectively.

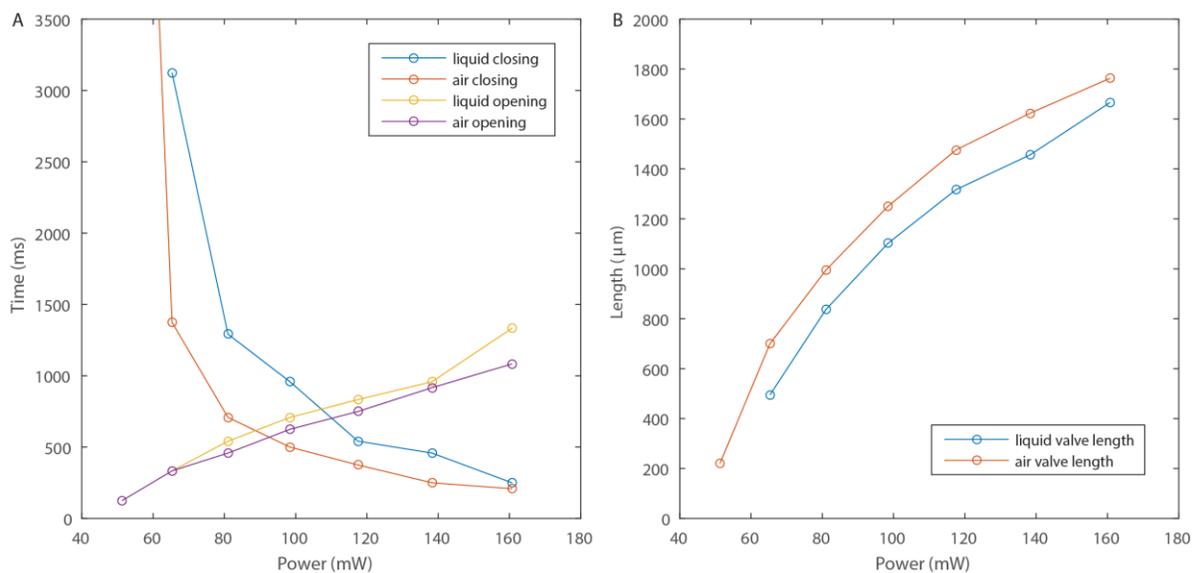

**Supplementary figure 8.** Liquid vs Air. A) Opening and closing times of the same valve with liquid and air. B) Length of the closed valve with liquid and air. PDMS thickness was 160 µm, channel height was 11.27 µm, and channel width was 100 µm.

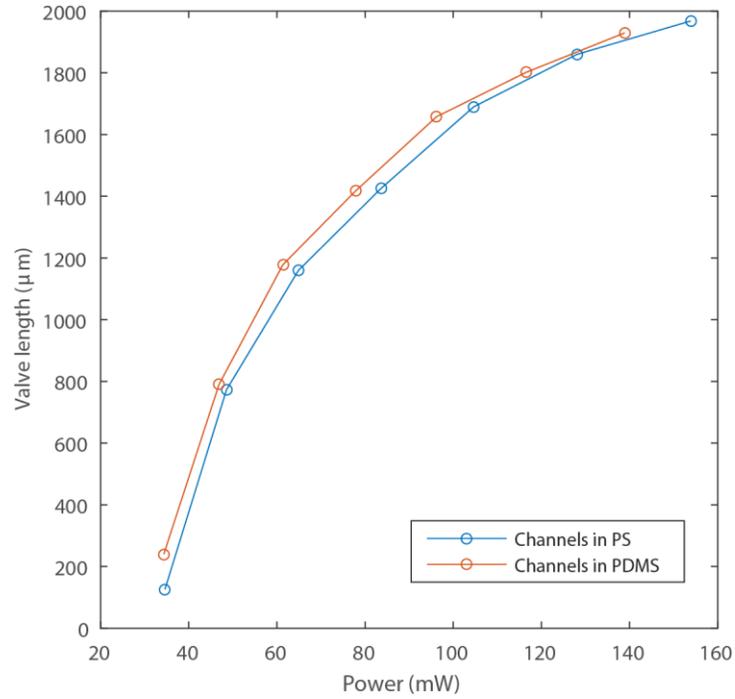

**Supplementary figure 9.** Alternative channel placement. Valve length vs power is comparable for different channel placements. The blue trace is from a chip with the channel placed in the PS layer, and the red trace is from a chip with the channel placed in the PDMS layer. For both devices, PDMS thickness was 160 µm, and channel width was 180 µm. For the device with channels in the PDMS, the channel height was 9.3 µm, and for the device with channels in the PS channel height was 11.1 µm.

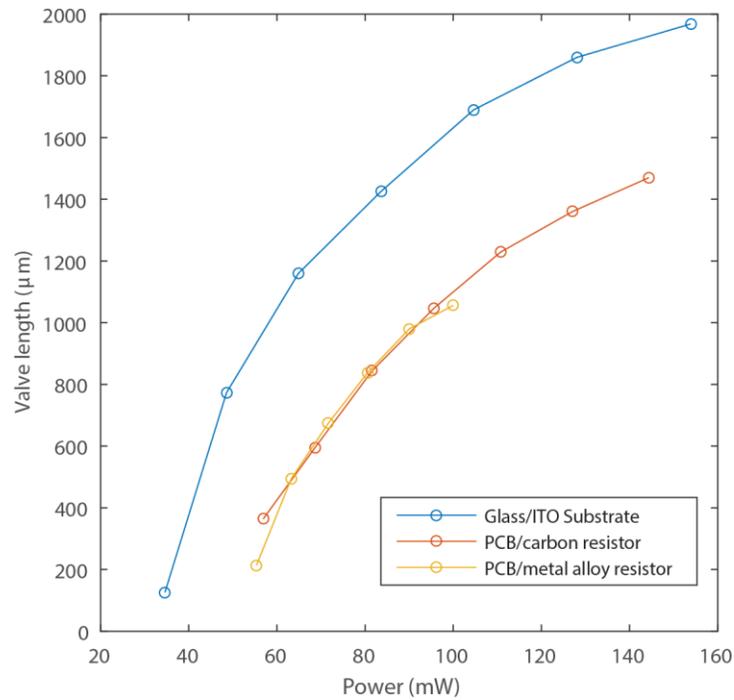

**Supplementary figure 10.** Alternative substrates. Valve length vs power depends on substrate. Blue trace is a glass base chip with ITO resistor, gold is a PCB base chip with metal alloy resistors, and red is PCB base chip with carbon resistors. For all devices, PDMS thickness was 160 µm, channel height was 11.1 µm, and channel width was 180 µm.

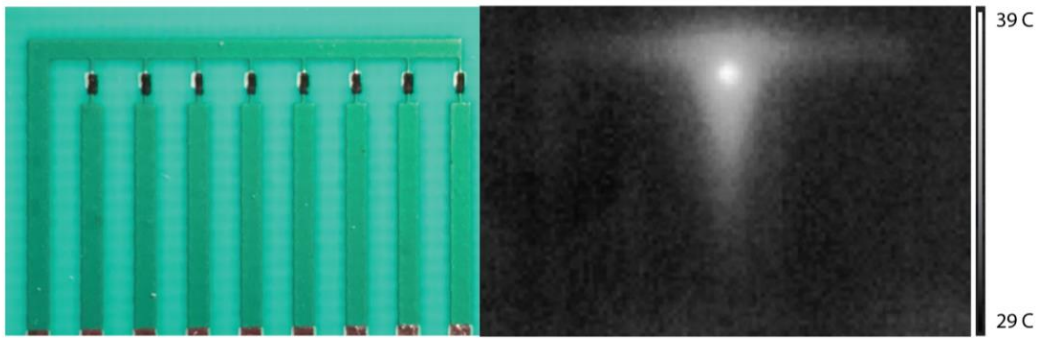

**Supplementary figure 11.** Heat loss from copper traces. Heat can be seen escaping the proximity of the valve along the copper traces. Visible light image of the PCB covered in PDMS on the left, thermal image on the right.

**Supplementary videos:**

1) An elastomeric focusing valve. Power was applied for one second while the red light was on, then removed for one second while the red light was off. PDMS layer thickness was 160 µm, channel height was 11.3 µm, and channel width was 100 µm.

2) Bubble formation upon valve opening. The valve was powered for 30 s, then power was removed, causing a bubble to form. PDMS layer thickness was 160 µm, channel height was 11.3 µm, and channel width was 100 µm.

3) Bubble formation was suppressed by 5 psi. The valve was powered for 30 s under 5 psi pressure, then power was removed while still under pressure. Bubble formation was inhibited. Channel dimensions are the same as in supplementary video 2.

4) Channels in PDMS layer. Power was applied for one second while the red light was on, then removed for one second while the red light was off. PDMS layer thickness was 160 µm, channel height was 11.03 µm, and channel width was 100 µm. The channel was in the PDMS layer, unlike the valve in supplementary video 1 where the channel was in the polystyrene layer.

5) An elastomeric focusing valve on a printed circuit board (PCB). The PCB is FR4 fiberglass with Nickel Chromium resistors (Ticer) and copper traces. Power was applied for one second while the red light was on, then removed for one second while the red light was off. PDMS layer thickness was 160 µm, channel height was 11.32 µm, and channel width was 140 µm.

6) An elastomeric focusing valve on a printed circuit board (PCB). The PCB is CEM-1 fiberglass with printed carbon resistors and copper traces. Power was applied for one second while the red light was on, then removed for one second while the red light was off. PDMS layer thickness was 160 µm, channel height was 11.4 µm, and channel width was 100 µm.

7) Light actuated elastomeric focusing valve. Absorption of a 100mW green laser beam by black PDMS created local heating and was used to actuate a valve with dynamic positioning by moving the device relative to the beam. Text shows FWHM of the beam. The video appears red from the filter blocking the laser light.

8) Light actuated elastomeric focusing valve in selective locations. Absorption of a 100mW green laser beam by selectively patterned black PDMS (squares) created local heating and actuated valves only when the beam contacted the black patterned regions. Text shows FWHM of the beam. The video appears red from the filter blocking the laser light.

9 Portable valve controller. A portable valve controller was created on the back of a smartphone case. An Arduino micro board was programmed to deliver on and off signals to a transistor at two second intervals. The transistor controlled whether or not power went from a 9 V battery to the valve. Pulse width modulation was used to tune the exact power level. A button turned the controller on and off, and we used as status indicators. The phone camera was used only to image the valve actuating.

**Citations**